# Transport Evidence for Wigner Crystals in Monolayer MoTe$_2$

Mingjie Zhang[1,2,†], Zhenyu Wang[1,2,†], Yifan Jiang[3,†], Yaotian Liu[1,2], Kenji Watanabe[4], Takashi Taniguchi[5], Song Liu[6], Shiming Lei[3,*], Yongqing Li[1,2,*], Yang Xu[1,2,*]

[1]Beijing National Laboratory for Condensed Matter Physics, Institute of Physics, Chinese Academy of Sciences, Beijing 100190, China

[2]School of Physical Sciences, University of Chinese Academy of Sciences, Beijing 100049, China

[3]Department of Physics, Hong Kong University of Science and Technology, Clear Water Bay, Hong Kong SAR, China

[4]Research Center for Functional Materials, National Institute for Materials Science, 1-1 Namiki, Tsukuba 305-0044, Japan

[5]International Center for Materials Nanoarchitectonics, National Institute for Materials Science, 1-1 Namiki, Tsukuba 305-0044, Japan

[6]Institute of Microelectronics, Chinese Academy of Science, Beijing 100029, China

† These authors contributed equally

* Email: phslei@ust.hk, yqli@iphy.ac.cn, yang.xu@iphy.ac.cn

**Abstract**

**The crystallization of charge carriers, dubbed the Wigner crystal, is anticipated at low densities in clean two-dimensional electronic systems (2DES). While there has been extensive investigation across diverse platforms, probing spontaneous charge and spin ordering is hindered by disorder effects and limited interaction energies. Here, we report transport evidence for Wigner crystals with antiferromagnetic exchange interactions in high-quality, hexagonal boron nitride encapsulated monolayer MoTe$_2$, a system that achieves a large interaction parameter ($r_s$) at proper hole densities. A density-tuned metal-insulator transition (MIT) occurring at $3.1\times10^{11}$ cm$^{-2}$ (corresponding to $r_s$~32) and pronounced nonlinear charge transport in the insulating regime at low temperatures signify the formation of Wigner crystals. Thermal melting of the crystalline phase is observed below approximately 2 K via temperature-dependent nonlinear transport. Magnetoresistance measurements further reveal a substantial enhancement of spin susceptibility as approaching the MIT. The temperature dependence of spin susceptibility in the Wigner crystal phase closely follows the Curie-Weiss law, with the extracted negative Weiss constant illustrating antiferromagnetic exchange interactions. Furthermore, we have found the system exhibits metallic-like differential resistivity under finite DC bias, possibly indicating the existence of a non-equilibrium coherent state in the depinning of Wigner crystals. Our observations establish monolayer MoTe$_2$ as a promising platform for exploring magnetic and dynamic properties of Wigner crystals.**

Main text

The correlated charge carriers confined in two dimensions serve as a prototypical platform for studying many-body physics. When the Coulomb interaction energy significantly exceeds the kinetic energy, the electronic properties of the system exhibit pronounced deviations from the anticipations of the Fermi liquid theory. To characterize the correlation strength of a 2D electronic system (2DES) with parabolic dispersions, a dimensionless parameter $r_s$ is defined as the ratio of interaction to kinetic energy, given by $r_s = m^*e^2/(4\pi\varepsilon\hbar^2\sqrt{\pi|n|})$, where $m^*$ is the effective mass of the charge carrier, $e$ is the elementary charge, $\varepsilon$ is the dielectric constant, $\hbar$ is the reduced Planck constant, and $n$ is the carrier density. In the strongly correlated regime, a clean 2DES is anticipated to form Wigner crystals (WCs) [1], a crystalline electron phase that spontaneously breaks the continuous translational symmetry, occurring at $r_s \geq \sim 30$ as shown by Monte Carlo (MC) calculations[2-4]. The WC-related phenomena have been extensively explored in conventional semiconductor heterostructures, including solidification of charge carriers at zero magnetic field[5-14], and in the quantum Hall regime under strong magnetic fields[15-27]. The metal-insulator transitions (MITs) at large $r_s$ values and the nonlinear transport characteristics are widely regarded as evidence for WCs. However, achieving strongly correlated charge carriers in these systems necessitates ultralow carrier densities, a regime where disorder effects such as Anderson localization emerge, obscuring the mechanisms underlying charge solidification[28,29]. For example, in systems such as $p$-GaAs or $n$-ZnO, the critical density for MIT at zero magnetic field falls below or about $1 \times 10^{10}$ cm$^{-2}$, where the average inter-charge separation is ~100 nm[5,8]. Furthermore, the corresponding ultralow exchange energy scale (< ~10 mK) sparks a controversy about the manifestations of magnetic states[7,8,30], which are predicted by MC calculations[2-4]. Beyond magnetic ordering, molecular dynamics (MD) simulations incorporating varying disorder strengths have identified diverse dynamical phases of moving Wigner solids[31-33], while dynamic properties associated with the depinning of WCs and the resulting nonlinear transport characteristics are still not fully understood.

Recent advances in atomically thin semiconducting transition metal dichalcogenides (TMDCs) have demonstrated their strong potential for probing the phase diagrams of correlated charge carriers at elevated densities comparing to conventional semiconductor heterostructures[12-14,34,35]. In electron-doped monolayer MoSe$_2$ systems, optical reflection experiments have revealed the signatures of WCs[12,14]. Further insights into the quantum melting of WCs have emerged from magneto-optical spectroscopy studies, which suggests the presence of an electronic microemulsion, a mixture of Fermi liquid and WCs, within the intermediate correlation regime[14]. This microemulsion picture gains additional experimental evidence from scanning tunneling microscopy studies conducted in the hole-doped bilayer MoSe$_2$ system[35]. Despite these advances, transport experiments investigating strongly correlated charge carriers in TMDCs face substantial challenges, primarily due to the inherent difficulty in fabricating ohmic contacts at low temperatures. While a charge-transfer contact scheme in monolayer WSe$_2$ system has enabled the observation of a MIT at $r_s = 26.4$[34], fundamental questions regarding the nature of the MIT, the magnetic ordering and the non-equilibrium dynamic behavior of WCs, remain unresolved. Systematic low-temperature transport measurements in the large $r_s$ regime are crucial to address these open questions. Here, we present transport studies in high-quality monolayer MoTe$_2$ devices featuring highly transparent contacts. Our observations demonstrate that WCs exhibit antiferromagnetic exchange interactions and provide insights into the non-equilibrium properties associated with their depinning. Meanwhile, as MoTe$_2$-based moiré superlattices have garnered significant attention in recent years for hosting integer and fractional quantum topological states[36-41], the intrinsic transport properties of monolayer MoTe$_2$ remain unexplored. Our findings fill this gap and open new avenues for studying strongly correlated electron systems in TMDCs based heterostructures.

The device structure of Sample A is illustrated schematically in Figure 1a. The hBN encapsulation provides an environment with a relatively low dielectric constant ($\varepsilon \approx 4.5\ \varepsilon_0$) for Coulomb interactions[42]. Together with the large effective mass of holes in monolayer MoTe$_2$ ($m^* \approx 0.75\ m_e$, where $m_e$ is the bare electron mass, Extended Data Fig. 3), $r_s$ reaches 30 at $n = -3.5 \times 10^{11}$ cm$^{-2}$. This value is nearly three times larger than that of monolayer WSe$_2$ and one order larger than that of conventional semiconductor heterostructures (such as AlAs or ZnO shown Figure 1b)[7,8,34]. The large $r_s$ at higher densities allows us to investigate the ground states of correlated charge carriers and their magnetic interactions over broader energy scales.

To optimize sample quality and achieve Ohmic contacts at low hole densities, two major improvements are implemented. First, the MoTe$_2$ monolayer is exfoliated from a high-quality flux-grown bulk crystal (see methods), yielding much lower disorders and higher mobilities compared with commercial crystals. Second, low contact resistance is achieved using local contact gates fabricated via a pre-patterning technique, which effectively mitigates potential electron-beam damage to the MoTe$_2$ flake (see Methods and Extended Data Fig. 1 for details). We note that all experimental data presented in the main text are collected from Sample A, unless otherwise specified. Figure 1c illustrates the temperature dependence of the mobility at different densities. For $|n| > \sim 2 \times 10^{12}$ cm$^{-2}$, the mobility follows a power-law dependence, $\mu \sim T^{-1}$, at high temperatures, suggesting acoustic phonon scattering above the Bloch-Grüneisen temperature[43]. At low temperatures, the mobility surpasses 10,000 cm$^2$ V$^{-1}$ s$^{-1}$ (one order higher than typical samples fabricated from commercial crystals, see more discussions in Methods) and exhibits a saturation trend, indicating that impurity scatterings and/or electron-electron scatterings dominate over phonon scattering. With decreasing the hole density, the sample approaches the MIT (to be discussed in Fig. 2) and the mobility decreases.

The combination of a high hole mobility and excellent contact transparency enables the measurement of longitudinal resistivity $\rho_{xx}$ across a wide range of hole densities $n$ and perpendicular magnetic fields $B$ at $T = 0.3$ K (see Extended Data Fig. 4). In Figure 1d, we convert the data into $\partial \rho_{xx}/\partial |n|$ and show the Landau fan diagram more clearly. All Landau levels (LLs) converge to the same band edge at $n = 0$ in the zero $B$ field limit. The detection of Shubnikov–de Haas (SdH) oscillations at magnetic fields as low as ~2 T (only show data above 4 T here), together with the identification of the first LL ($\nu_{LL}$ = 1), further confirms the high quality of the sample. In the high-density regime, the intensity of $\partial\rho_{xx}/\partial|n|$ exhibits a visible beating pattern for LLs. This behavior arises from an interplay between the orbital cyclotron energy $E_c = \hbar\omega_c$ (where $\omega_c = eB/m^*$ is the cyclotron frequency) and the density-dependent Zeeman energy $E_Z = 2g^*\mu_B B$ (where $g^*$ is the effective g-factor, $\mu_B$ is the Bohr magneton)[42,44-48]. The orange triangles mark the densities where the spin/valley degeneracy of LLs are lifted most evidently, corresponding to the condition $E_Z = (N+0.5)E_c$. From this equation, we can deduce the spin susceptibility $\chi = E_Z/E_c = g^*m^*/m_e = N + 0.5$ at these densities (see more details in Extended Data Fig. 5). As shown in Figure 1e, the extracted spin susceptibility exhibits a substantial increase with decreasing density, a phenomenon attributed to enhanced interaction effects at lower densities. Compared with other TMDCs (such as $WSe_2$, $MoSe_2$, and $MoS_2$ shown here), p-type monolayer $MoTe_2$ has the highest spin susceptibility, which is nearly two times larger at the same densities[42,44-48]. It is likely attributed to the combined effects of their large effective mass and the large orbital magnetic moment originating from the dominant orbital composition of the valence band maximum[45,49,50]. This large spin susceptibility, and the rapid rise at low densities, underscore the uniqueness of monolayer $MoTe_2$ as a platform for investigating strongly interacting holes.

In Figure 2a, we present $\rho_{xx}$ as a function of temperature at different densities. In the low temperature regime, the sample exhibits a metallic behavior ($d\rho_{xx}/dT > 0$) at higher densities, and changes to an insulating behavior ($d\rho_{xx}/dT < 0$) at lower densities. The curves reveal a MIT near the critical density $n_{MIT} \approx -3.1\times10^{11}$ cm$^{-2}$, corresponding to $r_s \approx 32$. The metallic behavior observed at high temperatures persists across a broad range of densities and temperatures, indicating that hole-phonon scattering is the dominant mechanism governing the temperature-dependent transport properties. Figure 2b displays the I-V characteristics of the sample near $n_{MIT}$ at T = 0.3 K (see Methods for measurement details). On the metallic side (where $|n| > |n_{MIT}|$), the sample exhibits linear I-V curves, with the differential resistivity $dV/dI$ remaining approximately constant as a function of $I_{DC}$. Conversely, on the insulating side (where $|n| < |n_{MIT}|$), the sample develops pronounced nonlinear I-V curves as the hole density decreases. The nonlinearity is characterized by a threshold voltage ($\Delta V$), that is extrapolated from the high-current quasi-linear portion of the I-V curve to zero current. Our data demonstrate that hBN encapsulated monolayer $MoTe_2$ is one of the few systems where a MIT is observed at $r_s > 30$ in the absence of a magnetic field (besides p-GaAs[5,51]). The nonlinear transport characteristics in the insulating phase could arise from trivial mechanisms, such as percolation or charge hopping driven by strong electric fields [52-55]. However, we note that prior optical reflection experiments in the monolayer $MoSe_2$ system[12,14], which have nearly identical effective masse ($m^*$) and dielectric constants ($\varepsilon$) with monolayer $MoTe_2$ system, reveal that charges prefer to form a crystalline structure when the carrier density falls below ~3×10$^{11}$ cm$^{-2}$. These facts strongly suggest the presence of WCs. Consequently, the reduction in differential resistivity at large DC bias arises from the depinning of WCs, with $\Delta V$ serving as a measure of the pinning strength.

To further probe the properties of WCs, we extend our experiment to the temperature-dependent nonlinear transport. Figure 3a shows the *I-V* traces measured at various temperatures for a density of $n \approx -3.3 \times 10^{11}$ cm$^{-2}$, near $n_{MIT}$ but slightly towards the metallic side for comparison. The weak $I_{DC}$ dependence of d$V$/d$I$ between 0.3 K and 2.2 K confirms the linear *I-V* characteristics. We note that the nonmonotonic temperature dependence of resistivity could be associated with the microemulsion phase[56], though the supporting evidence from transport measurements is indirect. In contrast, at $n \approx -2.2 \times 10^{11}$ cm$^{-2}$ on the insulating side (Figure 3b), the *I-V* trace is linear at $T = 2.2$ K, but gradually becomes nonlinear as the temperature decreases. Additionally, the full width at half maximum of the d$V$/d$I$ – $I_{DC}$ curves narrows with decreasing temperature. Figure 3c systematically displays the temperature dependence of d$V$/d$I$ at various $I_{DC}$ values. At zero DC bias, d$V$/d$I$ increases monotonically as the temperature is lowered. Intriguingly, under higher $I_{DC}$ conditions, the differential resistivity exhibits a metallic-like behavior at low temperatures, with d$V$/d$I$ peak shifting progressively to higher temperatures as $I_{DC}$ increases. The current-induced Joule heating effect can be safely excluded as it typically induces resistivity saturation at low temperatures, rather than the observed metallic-like trend. Overdamped MD simulations in systems with weak disorders have shown that pinned WCs can transit into a novel non-equilibrium phase of moving WCs under applied electric fields[32]. Experimentally, recent studies in GaAs quantum wells, have reported bias-induced negative, fluctuating differential resistances linked to a glass phase, where filamentary flow patterns may emerge under disordered conditions[27,31-33]. The anomalous behavior observed in our monolayer MoTe$_2$ sample implies that the depinned WCs induced by the DC bias became more mobile at lower temperatures, likely reflecting improved coherent transport. This could suggest the breaking of the continuous translational symmetry can still be preserved under the DC bias, providing preliminary evidence for the moving WCs. Further real-space microscopic and spectroscopic probes are warranted to confirm this interpretation[12-14,35,57].

We further quantify the strength of nonlinearity in the *I-V* curves by the ratio of zero DC bias differential resistivity ($\rho_0$) to that at a finite current of $I_{DC} \approx 5$ nA ($\rho_{5nA}$). As shown in Figure 3d, the nonlinearity is prominently observed in the insulating regime. We define the critical temperature $T_c$ as the point at which the charge transport exhibits a transition from nonlinear to linear behaviors. The dashed line in Figure 3d guides the evolution of $T_c$ as a function of carrier density. As the density decreases, $T_c$ exhibits a rapid increase near the $n_{MIT}$ and saturates at approximately 2 K at the lowest measured densities. The striking contrast between nonlinear and linear charge transport regimes highlights intrinsic differences in the states of holes. Above $T_c$, the disappearance of nonlinear charge transport signals the melting of WCs. The melting temperature (< 2 K) is comparable to the theoretically estimated value of 1 K[29,58,59], indicating disorder is relatively weak in our sample[29,58].

In addition to charge order, spin order is another long-sought phenomenon in the exploration of ground states for strongly interacting charge carriers[2-4,7,8,30]. We examine the spin susceptibility in the low-density regime through magnetoresistance measurements. Figure 4a presents the resistivity as a function of perpendicular magnetic field $B$ at various densities. All curves exhibit drastic positive magnetoresistance at low fields, with a weak field-dependent background superimposed on the SdH oscillations observed at higher magnetic fields. This large magnetoresistance and subsequent saturation are strongly reminiscent of the observation in low-density 2DES under in-plane magnetic fields, which drive the band to be fully spin polarized without introducing any orbital effects[7,8,56,60-67]. In our case, the in-plane magnetic field has almost no effects because of the large Ising spin-orbit coupling in monolayer TMDCs. With increasing the out-of-plane $B$ field or equivalently the Zeeman energy $E_z$ here, more holes will be transferred from one spin-polarized valley to the other (as depicted for $K$ and $K'$ valley in the inset of Figure 4b), resulting in net spin/valley polarization (charge imbalance between the two spin/valley species) while conserving the total charge density. In the polarization process (low field regime), the decreased Fermi momentum of spin-down holes weakens the screening of disorder potential seen by the majority spin-up holes, leading to a large positive magnetoresistance[68,69]. We note that this simplified interpretation ignores the magnetic-field-tuned interaction effects and the orbital effects. The system becomes fully spin (valley) polarized when $B$ reaches a critical magnetic field $B_c$, at which $E_z$ equals the Fermi energy ($E_F$). The critical field $B_c$ and its relationship with spin susceptibility $\chi$ hence satisfy the following equation:

$$B_c \approx \frac{\pi \hbar^2}{\mu_B} \frac{|n|}{g^* m^*} = \frac{\pi \hbar^2}{\mu_B} \frac{|n|}{m_e \chi}. \tag{1}$$

In the experiment, as shown in the topmost curve of Figure 4a, linear fits are applied to both sides of the inflection points in the log ($\rho_{xx}$) - $B$ curves. Subsequently, $B_c$ is determined as the intersection point of two fitted lines. For $|n| > 1.2 \times 10^{12}$ cm$^{-2}$, the linear fits are disrupted by SdH oscillations. Nevertheless, the critical field $B_c$ can be determined by tracing the evolution of the spin/valley degeneracy $D$ of LLs, where $D = \Phi_0 |n|/B_F$, $\Phi_0 = h/e$ is the magnetic flux quantum and $B_F$ is the SdH oscillation frequency. The precise $B_c$ lies within the range of magnetic fields associated with the two nearest spin/valley degenerate and spin/valley polarized LLs, and is estimated from the median value (Extended Data Fig. 6).

Figure 4b summarizes the density dependence of $B_c$ extracted from both SdH oscillations and magnetoresistances analyses. As we lower the hole density, $B_c$ decreases rapidly. We can estimate the spin susceptibility $\chi = g^* m^*/m_e = \pi \hbar^2 |n|/(\mu_B m_e B_c)$ and its inverse $\chi^{-1} \propto B_c$ using equation (1). As shown in Figure 4c, the values of $\chi$ derived from three distinct methods exhibit a consistent and smoothly varying trend, with the largest $\chi$ approaching ~29 and being over four times of the high-density value. This signals the possibility of observing long-range magnetic ordering in the low-density limit.

Temperature dependence of the spin susceptibility enables direct probing of the magnetic interactions between charge carriers. Figure 4d displays magnetoresistance at four different temperatures for $n = -2.8 \times 10^{11}$ cm$^{-2}$, corresponding to $r_s \approx 34$, where a WC forms. As depicted in Fig. 4e, the extracted critical field $B_c$ is plotted as a function of temperature $T$, showing a linear temperature dependence. This behavior implies that the spin susceptibility $\chi$ follows the Curie-Weiss law, expressed as $\chi^{-1} \propto T - \theta$, where $\theta$ denotes the Weiss constant and is extrapolated to $-0.9 \pm 0.1$ K in the zero-field limit. A negative $\theta$ typically signifies antiferromagnetic exchange interactions between localized magnetic moments. The antiferromagnetic interactions in the WCs possibly indicate that their ground states are antiferromagnetic, as predicted by MC calculations[4]. In contrast to monolayer MoSe$_2$ system, where the measured spin susceptibility of WCs follows a Curie law $\chi^{-1} \propto T$, suggesting the absence of spontaneous magnetism[14], the observed antiferromagnetic interactions in monolayer MoTe$_2$ is consistent with the larger spin susceptibilities. The hole density of WC in monolayer MoTe$_2$ is one order larger than that in traditional semiconductor heterostructures, corresponding to a Fermi temperature $E_F \approx 12$ K. The shorter distances between charge carriers generate stronger Coulomb interactions, leading to a larger exchange energy $|J|$, which is estimated to be on the order of 0.1 K through MC calculations[4,8]. The enhanced $E_F$ and $|J|$ in the monolayer MoTe$_2$ can facilitate the formation of magnetic WCs. The extracted $|\theta|$ value is on the same order as the estimated $|J|$. Moreover, theoretical studies propose that interstitial defects could introduce additional exchange interactions, potentially complicating the magnetic behavior of WCs in experiments[30]. The MC calculations also predict the emergence of ferromagnetic WCs in the strong correlation limit ($r_s > \sim 38$)[2-4]. In monolayer MoTe$_2$, the pronounced enhancement of $\chi$ indicate that the system may undergo a spontaneous spin/valley polarization at lower densities and temperatures; however, further evidence is required to confirm this interpretation.

Our experimental findings unveil the correlations among holes in the monolayer MoTe$_2$ system. The emergence of WCs at charge densities $\sim 3 \times 10^{11}$ cm$^{-2}$, coupled with significantly enhanced spin susceptibility, establishes this platform as a compelling candidate for further exploring WC physics. Additionally, our study highlights the system's great potential for investigating diverse correlated quantum phases, including antiferromagnetic and moving WCs. Future work could benefit from combined transport and scanning tunneling microscopy measurements performed under ultra-low temperature conditions[14,57]. This approach may enable direct characterization of the symmetry of moving WCs and provide critical insights into their collective dynamical properties.

1. Wigner, E. On the interaction of electrons in metals. *Physical Review* **46**, 1002-1011 (1934).
2. Tanatar, B. & Ceperley, D. M. Ground state of the two-dimensional electron gas. *Phys. Rev. B* **39**, 5005-5016 (1989).
3. Attaccalite, C., Moroni, S., Gori-Giorgi, P. & Bachelet, G. B. Correlation energy and spin polarization in the 2D electron gas. *Phys. Rev. Lett.* **88**, 256601 (2002).
4. Drummond, N. D. & Needs, R. J. Phase diagram of the low-density two-dimensional homogeneous electron gas. *Phys. Rev. Lett.* **102**, 126402 (2009).
5. Yoon, J., Li, C. C., Shahar, D., Tsui, D. C. & Shayegan, M. Wigner crystallization and metal-insulator transition of two-dimensional holes in GaAs at B=0. *Phys. Rev. Lett.* **82**, 1744-1747 (1999).
6. Brussarski, P., Li, S., Kravchenko, S., Shashkin, A. & Sarachik, M. Transport evidence for a sliding two-dimensional quantum electron solid. *Nat. Commun.* **9**, 3803 (2018).
7. Hossain, M. S. *et al.* Observation of spontaneous ferromagnetism in a two-dimensional electron system. *Proc. Natl. Acad. Sci.* **117**, 32244-32250 (2020).
8. Falson, J. *et al.* Competing correlated states around the zero-field Wigner crystallization transition of electrons in two dimensions. *Nat. Mater.* **21**, 311–316 (2022).
9. Shashkin, A. & Kravchenko, S. Metal-insulator transition and low-density phases in a strongly-interacting two-dimensional electron system. *Ann. Phys.* **435**, 168542 (2021).
10. Melnikov, M., Shashkin, A., Huang, S., Liu, C. & Kravchenko, S. Collective depinning and sliding of a quantum Wigner solid in a two-dimensional electron system. *Phys. Rev. B* **109**, L041114 (2024).
11. Munyan, S., Ahadi, S., Guo, B., Rashidi, A. & Stemmer, S. Evidence of Zero-Field Wigner Solids in Ultrathin Films of Cadmium Arsenide. *Phys. Rev. X* **14**, 041037 (2024).
12. Smoleński, T. *et al.* Signatures of Wigner crystal of electrons in a monolayer semiconductor. *Nature* **595**, 53-57 (2021).
13. Zhou, Y. *et al.* Bilayer Wigner crystals in a transition metal dichalcogenide heterostructure. *Nature* **595**, 48-52 (2021).
14. Sung, J. *et al.* An electronic microemulsion phase emerging from a quantum crystal-to-liquid transition. *Nat. Phys.* **21**, 437-443 (2025).
15. Andrei, E. *et al.* Observation of a Magnetically Induced Wigner Solid. *Phys. Rev. Lett.* **60**, 2765-2768 (1988).
16. Goldman, V. J., Santos, M., Shayegan, M. & Cunningham, J. E. Evidence for two-dimentional quantum Wigner crystal. *Phys. Rev. Lett.* **65**, 2189-2192 (1990).
17. Williams, F. I. B. *et al.* Conduction Threshold and Pinning Frequency of Magnetically Induced Wigner Solid. *Phys. Rev. Lett.* **66**, 3285-3288 (1991).
18. Chen, Y. *et al.* Melting of a 2D quantum electron solid in high magnetic field. *Nat. Phys.* **2**, 452-455 (2006).
19. Csáthy, G., Tsui, D., Pfeiffer, L. & West, K. Astability and negative differential resistance of the Wigner solid. *Phys. Rev. Lett.* **98**, 066805 (2007).
20. Jang, J., Hunt, B. M., Pfeiffer, L. N., West, K. W. & Ashoori, R. C. Sharp tunnelling resonance from the vibrations of an electronic Wigner crystal. *Nat. Phys.* **13**, 340-344 (2017).
21. Knighton, T. *et al.* Evidence of two-stage melting of Wigner solids. *Phys. Rev. B* **97** (2018).
22. Deng, H. *et al.* Probing the Melting of a Two-Dimensional Quantum Wigner Crystal via its

Screening Efficiency. *Phys. Rev. Lett.* **122**, 116601 (2019).

23  Ma, M. *et al.* Thermal and Quantum Melting Phase Diagrams for a Magnetic-Field-Induced Wigner Solid. *Phys. Rev. Lett.* **125**, 036601 (2020).

24  Hossain, M. S. *et al.* Anisotropic Two-Dimensional Disordered Wigner Solid. *Phys. Rev. Lett.* **129**, 036601 (2022).

25  Zhao, L. *et al.* Dynamic Response of Wigner Crystals. *Phys. Rev. Lett.* **130**, 246401 (2023).

26  Tsui, Y.-C. *et al.* Direct observation of a magnetic-field-induced Wigner crystal. *Nature* **628**, 287-292 (2024).

27  Madathil, P. *et al.* Moving Crystal Phases of a Quantum Wigner Solid in an Ultra-High-Quality 2D Electron System. *Phys. Rev. Lett.* **131**, 236501 (2023).

28  Ahn, S. & Das Sarma, S. Density-tuned effective metal-insulator transitions in two-dimensional semiconductor layers: Anderson localization or Wigner crystallization. *Phys. Rev. B* **107**, 195435 (2023).

29  Vu, D. & Das Sarma, S. Thermal melting of a quantum electron solid in the presence of strong disorder: Anderson localization versus the Wigner crystal. *Phys. Rev. B* **106**, L121103 (2022).

30  Kim, K.-S., Murthy, C., Pandey, A. & Kivelson, S. A. Interstitial-induced ferromagnetism in a two-dimensional Wigner crystal. *Phys. Rev. Lett.* **129**, 227202 (2022).

31  Reichhardt, C., Olson, C., Gronbech-Jensen, N. & Nori, F. Moving Wigner glasses and smectics: Dynamics of disordered Wigner crystals. *Phys. Rev. Lett.* **86**, 4354-4357 (2001).

32  Reichhardt, C. & Reichhardt, C. Nonlinear dynamics, avalanches, and noise for driven Wigner crystals. *Phys. Rev. B* **106**, 235417 (2022).

33  Reichhardt, C. & Reichhardt, C. Noise and thermal depinning of Wigner crystals. *J. Condens. Matter Phys* **35**, 325603 (2023).

34  Pack, J. *et al.* Charge-transfer contacts for the measurement of correlated states in high-mobility $WSe_2$. *Nat. Nanotechnol.* **19**, 948-954 (2024).

35  Xiang, Z. *et al.* Quantum melting of a disordered wigner solid. Preprint at arxiv.org/abs/2402.05456 (2024).

36  Li, T. *et al.* Quantum anomalous Hall effect from intertwined moiré bands. *Nature* **600**, 641-646 (2021).

37  Cai, J. *et al.* Signatures of fractional quantum anomalous Hall states in twisted $MoTe_2$. *Nature* **622**, 63-68 (2023).

38  Zeng, Y. *et al.* Thermodynamic evidence of fractional Chern insulator in moiré $MoTe_2$. *Nature* **622**, 69-73 (2023).

39  Park, H. *et al.* Observation of fractionally quantized anomalous Hall effect. *Nature* **622**, 74-79 (2023).

40  Xu, F. *et al.* Observation of integer and fractional quantum anomalous Hall effects in twisted bilayer $MoTe_2$. *Phys. Rev. X* **13**, 031037 (2023).

41  Kang, K. *et al.* Evidence of the fractional quantum spin Hall effect in moiré $MoTe_2$. *Nature* (2024).

42  Larentis, S. *et al.* Large effective mass and interaction-enhanced Zeeman splitting of K-valley electrons in $MoSe_2$. *Phys. Rev. B* **97**, 201407 (2018).

43  Kaasbjerg, K., Thygesen, K. & Jacobsen, K. Phonon-limited mobility in n-type single-layer $MoS2$ from first principles. *Phys. Rev. B* **85**, 115317 (2012).

44      Movva, H. C. P. *et al.* Density-dependent quantum Hall states and Zeeman splitting in monolayer and bilayer WSe$_2$. *Phys. Rev. Lett.* **118**, 247701 (2017).

45      Gustafsson, M. V. *et al.* Ambipolar Landau levels and strong band-selective carrier interactions in monolayer WSe$_2$. *Nat. Mater.* **17**, 411-415 (2018).

46      Pisoni, R. *et al.* Interactions and magnetotransport through spin-valley coupled Landau levels in monolayer MoS$_2$. *Phys. Rev. Lett.* **121**, 247701 (2018).

47      Shi, Q. *et al.* Odd- and even-denominator fractional quantum Hall states in monolayer WSe$_2$. *Nat. Nanotechnol.* **15**, 569-573 (2020).

48      Foutty, B. A. *et al.* Anomalous Landau Level Gaps Near Magnetic Transitions in Monolayer WSe$_2$. *Phys. Rev. X* **14**, 031018 (2024).

49      Zhu, Z. Y., Cheng, Y. C. & Schwingenschlögl, U. Giant spin-orbit-induced spin splitting in two-dimensional transition-metal dichalcogenide semiconductors. *Phys. Rev. B* **84**, 153402 (2011).

50      Kosmider, K., González, J. & Fernández-Rossier, J. Large spin splitting in the conduction band of transition metal dichalcogenide monolayers. *Phys. Rev. B* **88**, 245436 (2013).

51      Manfra, M. *et al.* Transport and percolation in a low-density high-mobility two-dimensional hole system. *Phys. Rev. Lett.* **99**, 236402 (2007).

52      Jiang, H. W., Stormer, H. L., Tsui, D. C., Pfeiffer, L. N. & West, K. W. Magnetotransport studies of the insulating phase around $\nu = 1/5$ Landau-level filling. *Phys. Rev. B* **44**, 8107-8114 (1991).

53      Dolgopolov, V. T., Kravchenko, G. V., Shashkin, A. A. & Kravchenko, S. V. Metal-insulator transition in Si inversion layers in the extreme quantum limit. *Phys. Rev. B* **46**, 13303-13308 (1992).

54      Shashkin, A. A., Dolgopolov, V. T. & Kravchenko, G. V. Insulating phases in a two-dimensional electron system of high-mobility Si MOSFET's. *Phys. Rev. B* **49**, 14486-14495 (1994).

55      Marianer, S. & Shklovskii, B. I. Effective temperature of hopping electrons in a strong electric field. *Phys. Rev. B* **46**, 13100-13103 (1992).

56      Spivak, B., Kravchenko, S. V., Kivelson, S. A. & Gao, X. P. A. Colloquium: Transport in strongly correlated two dimensional electron fluids. *Rev. Mod. Phys.* **82**, 1743-1766 (2010).

57      Li, H. *et al.* Imaging two-dimensional generalized Wigner crystals. *Nature* **597**, 650-654 (2021).

58      Huang, Y. & Das Sarma, S. Electronic transport, metal-insulator transition, and Wigner crystallization in transition metal dichalcogenide monolayers. *Phys. Rev. B* **109**, 245431 (2024).

59      Hwang, W. & Das Sarma, S. Plasmon dispersion in dilute two-dimensional electron systems: Quantum-classical and Wigner crystal-electron liquid crossover. *Phys. Rev. B* **64**, 165409 (2001).

60      Simonian, D., Kravchenko, S. V., Sarachik, M. P. & Pudalov, V. M. Magnetic field suppression of the conducting phase in two dimensions. *Phys. Rev. Lett.* **79**, 2304-2307 (1997).

61      Mertes, K. M., Simonian, D., Sarachik, M. P., Kravchenko, S. V. & Klapwijk, T. M. Response to parallel magnetic field of a dilute two-dimensional electron system across the metal-insulator transition. *Phys. Rev. B* **60**, R5093-R5096 (1999).

62      Okamoto, T., Hosoya, K., Kawaji, S. & Yagi, A. Spin degree of freedom in a two-dimensional electron liquid. *Phys. Rev. Lett.* **82**, 3875-3878 (1999).

63      Yoon, J., Li, C. C., Shahar, D., Tsui, D. C. & Shayegan, M. Parallel magnetic field induced transition in transport in the dilute two-dimensional hole system in GaAs. *Phys. Rev. Lett.* **84**, 4421-4424 (2000).

64      Shashkin, A. A., Kravchenko, S. V., Dolgopolov, V. T. & Klapwijk, T. M. Indication of the

| | ferromagnetic instability in a dilute two-dimensional electron system. *Phys. Rev. Lett.* **87**, 086801 (2001). |
|---|---|
| 65 | Tutuc, E., De Poortere, E. P., Papadakis, S. J. & Shayegan, M. In-plane magnetic field-induced spin polarization and transition to insulating behavior in two-dimensional hole systems. *Phys. Rev. Lett.* **86**, 2858-2861 (2001). |
| 66 | Vitkalov, S. A., Zheng, H., Mertes, K. M., Sarachik, M. P. & Klapwijk, T. M. Scaling of the magnetoconductivity of silicon MOSFETs: evidence for a quantum phase transition in two dimensions. *Phys. Rev. Lett.* **87**, 086401 (2001). |
| 67 | Li, S., Zhang, Q., Ghaemi, P. & Sarachik, M. P. Evidence for mixed phases and percolation at the metal-insulator transition in two dimensions. *Phys. Rev. B* **99**, 155302 (2019). |
| 68 | Dolgopolov, V. T. & Gold, A. Magnetoresistance of a two-dimensional electron gas in a parallel magnetic field. *JETP Lett.* **71**, 27-30 (2000). |
| 69 | Herbut, I. F. The effect of parallel magnetic field on the Boltzmann conductivity and the Hall coefficient of a disordered two-dimensional Fermi liquid. *Phys. Rev. B* **63**, 113102 (2001). |

**Acknowledgements**

We thank X. Dai, Y. Zhou, and Y. Yang for helpful discussions. This work was supported by the National Key R&D Program of China (Grant Nos. 2021YFA1401300 and 2022YFA1403403), the National Natural Science Foundation of China (Grant No. 12174439), and the Innovation Program for Quantum Science and Technology (Grant Nos. 2021ZD0302400 and 2021ZD0302300). S. Lei acknowledge support by the Hong Kong RGC (No. 26308524), the Hong Kong Collaborative Research Fund (No. C6053-23G), and the Ministry of Science and Technology of the People's Republic of China (No. MOST23SC01). The growth of hBN crystals was supported by the Elemental Strategy Initiative of MEXT, Japan, and CREST (JPMJCR15F3), JST.

**Author contributions**

Y.X. and M.Z. conceived the project. M.Z., Z.W., and T.L. fabricated the devices. M.Z. performed the measurements and analyzed the data. Y.J., S.Liu, and S.Lei grew the bulk $MoTe_2$ crystals, and K.W. and T.T. grew the bulk hBN crystals. M.Z. and Y.X. co-wrote the manuscript. All authors discussed the results and commented on the manuscript.

**Methods**

**Crystal growth**

High-quality 2H-MoTe$_2$ single crystals were grown using a self-flux method with an excess of tellurium (Te). High-purity molybdenum (Mo) powders (99.997%) and Te (99.999%) were mixed at a molar ratio of 1:100 (Mo:Te) and sealed in an evacuated quartz ampoule. The mixture was heated to 1000 °C over 12 hours, then slowly cooled to 540 °C at a rate of 2 °C per hour. The crystals were then separated from the flux by decanting in a centrifuge. The resulting crystals were separated from the flux by decanting in a centrifuge. To remove residual Te, the crystals were subsequently annealed under a temperature gradient ($T_{\text{hot}}$ = 400 °C, $\Delta T \approx$ 200 °C) for two days before being cooled to room temperature.

**Device fabrication**

The monolayer MoTe$_2$ devices were fabricated using the layer-by-layer dry-transfer technique. All 2D material flakes (MoTe$_2$, hBN, and graphite) were mechanically exfoliated from the bulk crystals. The fabrication process commenced with the transfer of a bottom gate structure (an hBN/graphite stack) onto a silicon substrate using a polycarbonate (PC) stamp. The Pt (7 nm) contact electrodes were deposited onto the hBN layer through electron-beam lithography followed by electron-beam evaporation. To connect the Pt electrodes and the external peripheral conductive pads, we deposited Ti/Au (5/25 nm) by another step of lithography and metallization. The central region of the Pt electrodes was cleaned using contact-mode atomic force microscope.

For Sample A, bulk MoTe$_2$ crystals were synthesized using an optimized self-flux growth method (see above). To mitigate degradation of the exfoliated monolayer and achieve ultrahigh contact transparency, a pre-patterned contact gate method was developed, as illustrated in the following steps. First, a hBN contact gate layer (~ 5 nm) was mechanically exfoliated onto a SiO$_2$/Si substrate. Subsequently, Ti/Au (2/5 nm) contact gate metal layers with a predefined geometry were deposited onto the hBN via electron-beam evaporation (Extended Data Figure 1). A PC stamp was then employed to pick up the top gate stack and subsequently a MoTe$_2$ monolayer within a glovebox environment maintaining oxygen and moisture levels below 1 ppm. The assembled heterostructure was precisely aligned and transferred onto pre-fabricated bottom Pt electrodes. Residual polymer contaminants were removed by immersion in chloroform for 10 minutes. To ensure electrical connectivity between the contact gate metal and external electrodes, an additional PC stamp was utilized to transfer a thick graphite flake as a bridging component. Finally, the completed device underwent a second chloroform immersion to eliminate remaining polymer residues.

Sample B was fabricated with a conventional dual-gate structure, where the MoTe$_2$ bulk crystals were commercially sourced from HQ Graphene. Following the fabrication of the bottom-gate structure, another PC stamp was used to pick up a graphite/hBN stack, which served as the top gate. We then use this stack to pick up the exfoliated MoTe$_2$ monolayer in the glovebox. The whole stack was released onto the prepared bottom Pt electrodes, followed by a 10-minute chloroform immersion to remove residual polymer contaminants. We have fabricated many devices using MoTe$_2$ bulk crystals procured from HQ Graphene. Among these, sample B exhibited the highest quality, with a mobility estimated to be less than 1,000 cm$^2$ V$^{-1}$ s$^{-1}$ at $n \approx -3\times10^{12}$ cm$^{-2}$. A metal-insulator transition was observed near the critical density of $n_{\text{MIT}} \approx -1.1\times10^{12}$ cm$^{-2}$, approximately three times higher than that of sample A, suggesting the presence of stronger disorder effects. The transport characterization of sample B are presented in Extended Data Figure 7 and Figure 8.

**Transport measurements**

All transport measurements were performed in a He-3 cryostat with a base temperature of $T \approx 300$ mK and an applied magnetic field $B$ of up to 9 T. The four-terminal longitudinal resistivity $\rho_{xx}$ was measured using a standard low-frequency lock-in technique (frequency $f$ = 8–15 Hz), with the experimental configuration schematically depicted in Extended Data Figure 2a. To ensure measurement accuracy, voltage preamplifiers (DL 1201) featuring a high input impedance (100 MΩ) were utilized. The setup for measuring $I$-$V$ characteristics, as illustrated in Extended Data Figure 2b, employs a DC voltage source (Yokogawa 7651) and an AC voltage source (NF LI 5640 Lock-in Amplifier) connected in series to supply the input voltage. A 100 MΩ resistor is placed in series with the sample to limit the current magnitude. For differential resistance measurements, a small AC excitation current (0.1 nA) was applied to the sample. The output AC voltage $V_{\text{AC}}^{\text{OUT}}$ was measured using a DL 1201 voltage preamplifier and a Lock-in Amplifier, while a current preamplifier (SR 570) enabled simultaneous monitoring of the AC and DC output currents ($I_{\text{AC}}^{\text{OUT}}$ and $I_{\text{DC}}^{\text{OUT}}$, respectively). The differential resistance d$V$/d$I$ (calculated as $V_{\text{AC}}^{\text{OUT}}/I_{\text{AC}}^{\text{OUT}}$) was determined as a function of $I_{\text{DC}}$. Subsequently, the $V_{\text{DC}}$-$I_{\text{DC}}$ curves were derived by numerically integrating the d$V$/d$I$ – $I_{\text{DC}}$ datasets.

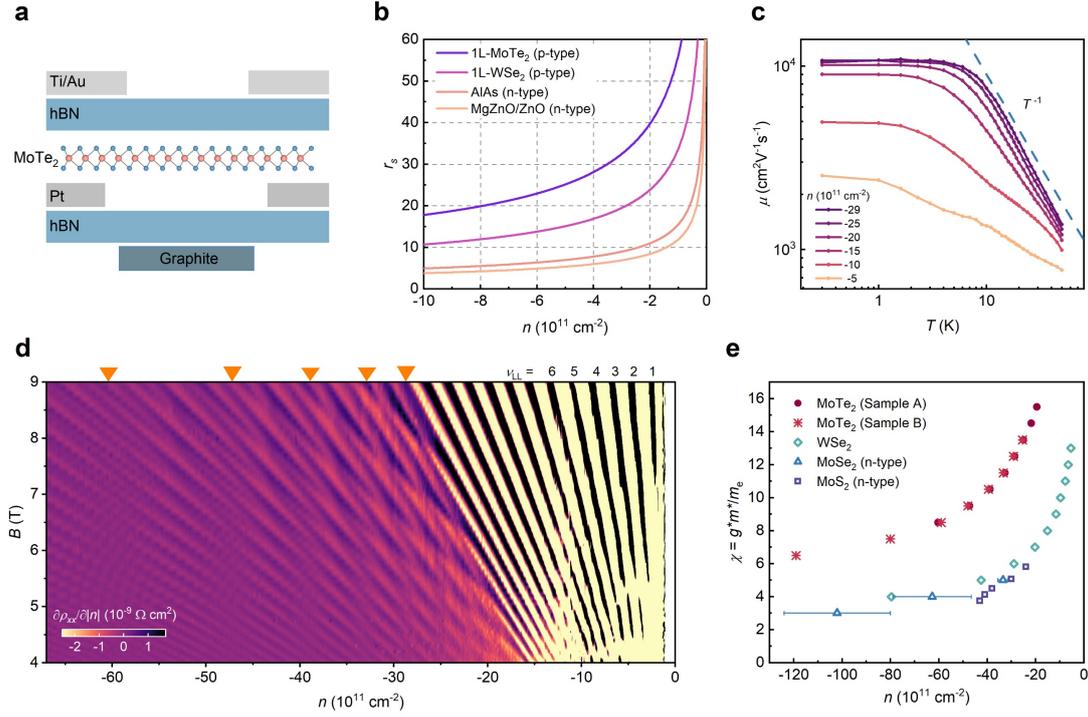

**Fig. 1 | Device structure and transport characterization. a**, Schematic of the device. The monolayer MoTe$_2$ is encapsulated by hBN, with pre-patterned Ti/Au pads serving as contact gates. **b**, The dimensionless parameter $r_s$ as a function of carrier density for hole doped monolayer MoTe$_2$, WSe$_2$, and electron doped AlAs and ZnO. **c**, Temperature dependence of mobility at various hole densities. Dashed line shows $\mu \propto T^{-1}$ as a guide. **d**, Mapping of $\partial \rho_{xx}/\partial |n|$ as a function of the magnetic field $B$ and the density $n$ at $T = 0.3$ K. The orange triangles mark the densities where the spin/valley degeneracy of LLs are lifted most evidently. **e**, Spin susceptibility $\chi = g^{*}m^{*}/m_e$ as a function of density. The spin susceptibilities of MoTe$_2$ determined in this work are plotted in red. For comparison, the reported spin susceptibilities for hole-doped monolayer WSe$_2$, electron-doped monolayer MoSe$_2$ and MoS$_2$ are also plotted[42,46-48].

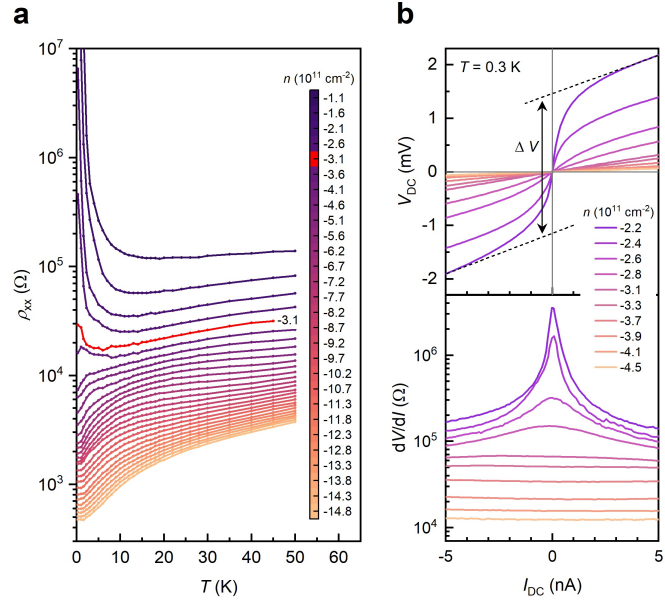

**Fig. 2 | Density-tuned metal-insulator transition and nonlinear transport characteristics. a**, Temperature dependence of the longitudinal resistivity $\rho_{xx}$ at various hole densities, showing the density-tuned metal-insulator transition. The red curve highlights the critical density $n_{MIT} \approx -3.1 \times 10^{11}$ cm$^{-2}$. **b**, Density dependent *I-V* curves (upper panel) and the corresponding differential resistivity d$V$/d$I$ (lower panel) at $T = 0.3$ K. The nonlinear transport characteristics emerge when $|n| < |n_{MIT}|$. The dashed lines are the extrapolations of the high-current portion of the *I-V* curve at $n \approx -2.2 \times 10^{11}$ cm$^{-2}$. The $\Delta V$ is the voltage drop between two dashed lines at zero $I_{DC}$.

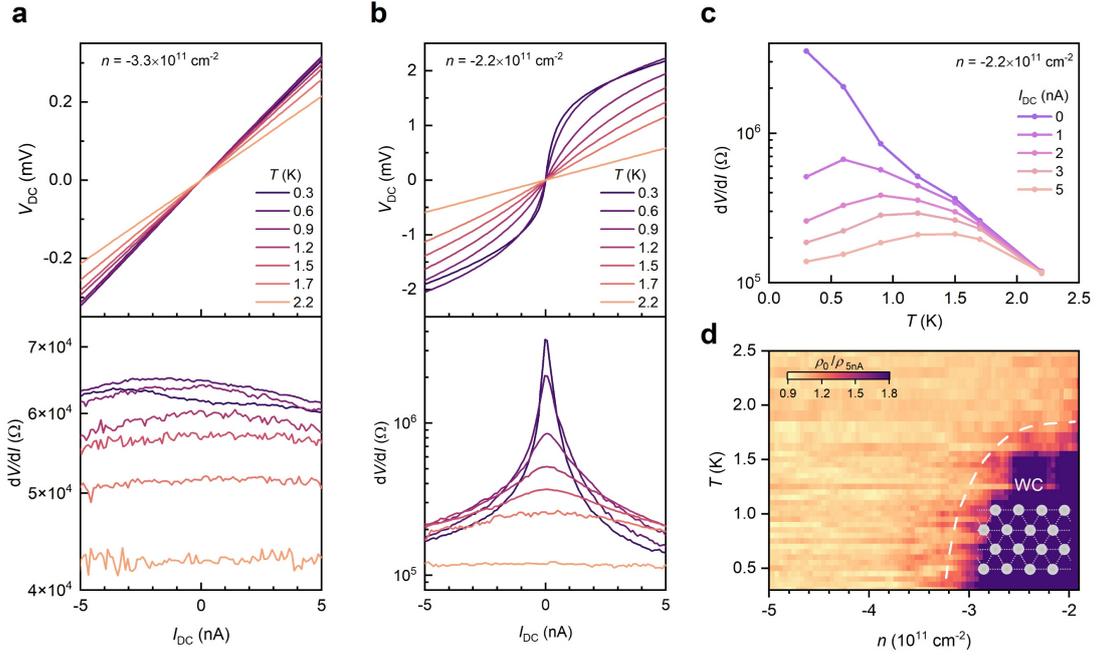

**Fig. 3 | Temperature dependent nonlinear transport characteristics. a**, Temperature dependent *I-V* curves (upper panel) and the corresponding differential resistivity (lower panel) at $n_{MIT} \approx -3.3 \times 10^{11}$ cm$^{-2}$. The *I-V* curves remain linear across this temperature range. **b**, Temperature dependent *I-V* curves (upper panel) and the corresponding differential resistivity (lower panel) at $n \approx -2.2 \times 10^{11}$ cm$^{-2}$. The nonlinear transport behavior disappears at $T = 2.2$ K. **c**, Differential resistivity d*V*/d*I* derived from **b** as a function of temperatures at various DC current values $I_{DC}$. **d**, Strength of nonlinearity $\rho_0/\rho_{5nA}$ in the *I-V* curves plotted as a function of *n* and *T*, where $\rho_0/\rho_{5nA}$ is the ratio of zero DC bias differential resistivity ($\rho_0$) to that at a finite current of $I_{dc} \approx 5$ nA ($\rho_{5nA}$). Dashed line guides the boundary of nonlinear transport regime. The inset displays the schematic diagram of a WC.

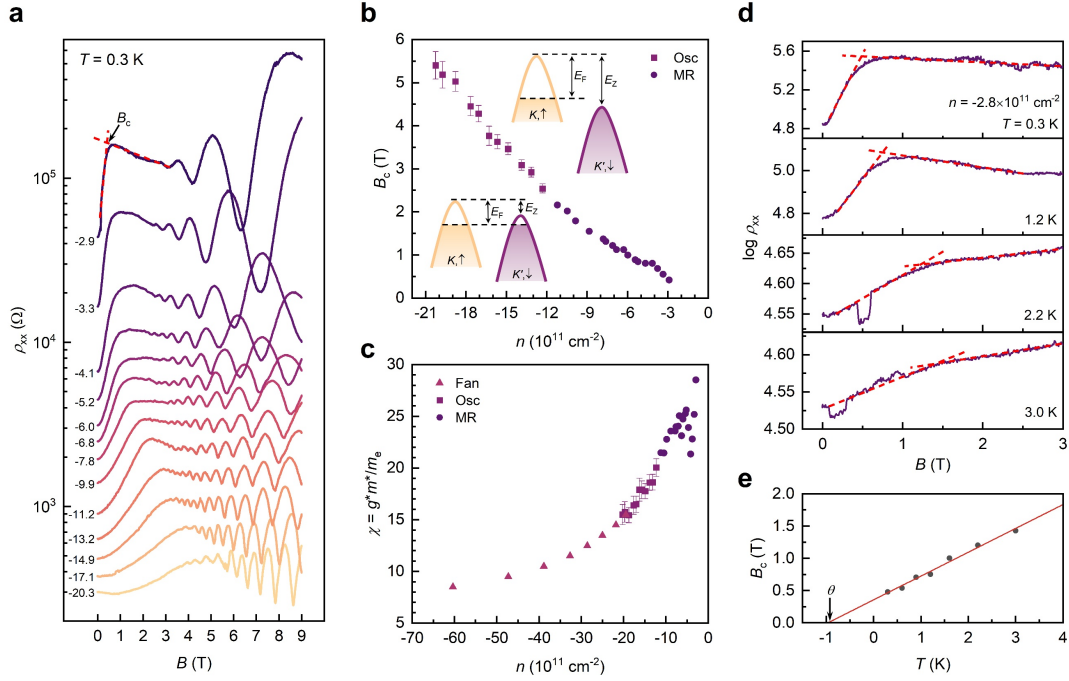

**Fig. 4 | Spin polarization and spin susceptibility. a**, The longitudinal resistivity $\rho_{xx}$ as a function of the perpendicular magnetic field $B$ at various densities. Each curve is labeled with the corresponding density in units of $10^{11}$ cm$^{-2}$. Two red dashed lines superimposed on the topmost $\rho_{xx}$-$B$ curve are the linear fits applied to both sides of the inflection points. The intersection point of two fitted lines denotes the critical magnetic field $B_c$ of full spin polarization. **b**, Density dependence of $B_c$ at $T = 0.3$ K. Squares and dots represent the $B_c$ extracted from SdH oscillations and magnetoresistances analyses, respectively. Lower and upper insets schematically illustrate the band alignments of the $K$ and $K'$ valleys for $B < B_c$ and $B > B_c$, respectively. For simplicity, the LLs are not drawn here. When the Zeeman energy $E_Z$ starts to exceed the Fermi energy $E_F$ at $B = B_c$, all carriers are polarized to one spin/valley. **c**, Density dependence of the spin susceptibility extracted from three different methods at $T = 0.3$ K. Triangles represent the measurements from Landau fan. Squares and dots represent the same measurements shown in **b**. **d**, The evolution of log $\rho_{xx}$ vs $B$ with temperature at $n \approx -2.8 \times 10^{11}$ cm$^{-2}$. **e**, Temperature dependence of $B_c$ ($\propto \chi^{-1}$) extracted from **d**. The red line is Curie-Weiss (linear) fit, with the intercept on the $T$-axis, denoted by $\theta$, indicating the Weiss constant.

**Extended Data Figures**

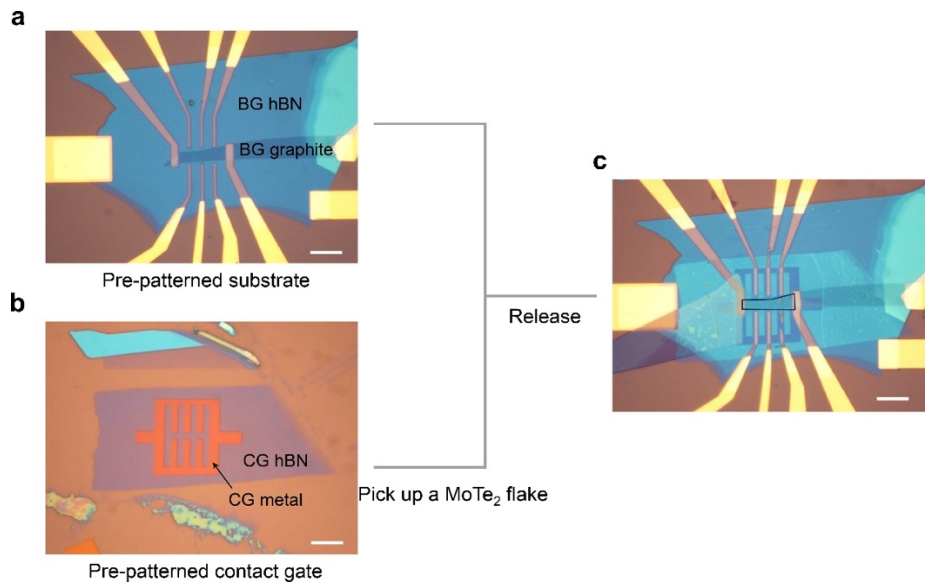

**Extended Data Figure 1 | Fabrication of Sample A. a**, Optical micrograph of the pre-patterned substrate, where the central contact electrodes consist of 7 nm-thick Pt. Graphite and hBN serve as the bottom gate (BG). **b**, Pre-patterned contact gate (CG). The geometry of the CG metal (Ti/Au 2/5 nm) is predefined to align with the Pt electrodes. **c**, The complete device. The conduction channel is well defined by the BG graphite, and outlined by the black polygon. The scale bars are 10 μm.

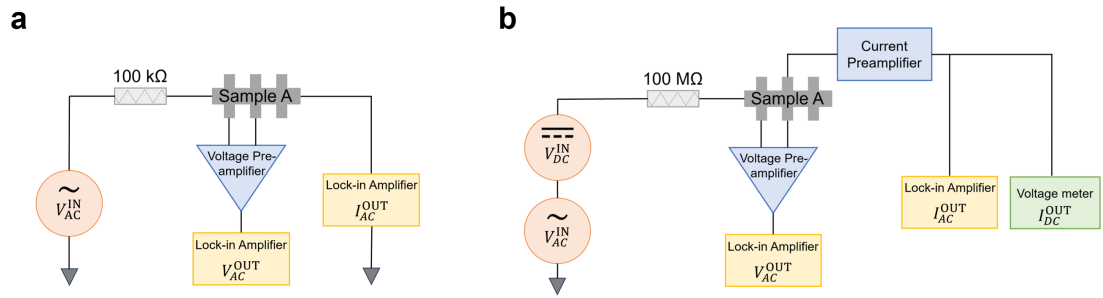

**Extended Data Figure 2 | Measurement setup. a**, Schematic of the measurement setup used for longitudinal resistance measurements via the low-frequency lock-in technique. **b**, Setup for measuring differential resistance as a function of DC current. Different source-drain electrodes are used to achieve better contact transparency when we measure the differential resistance.

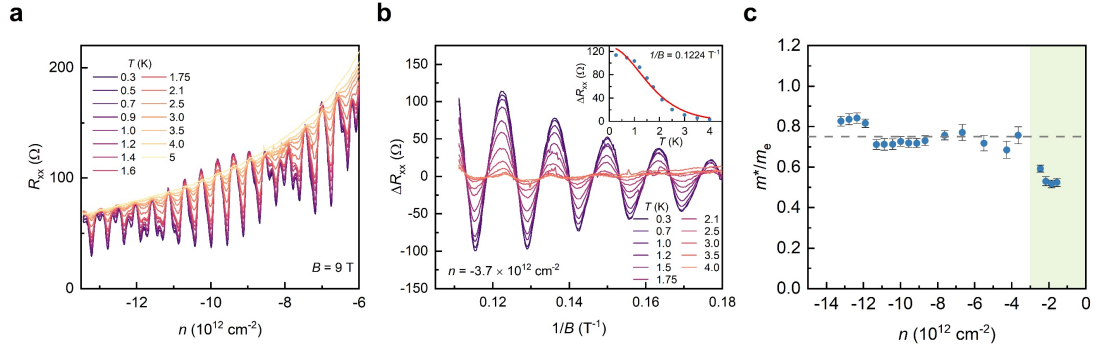

**Extended Data Figure 3 | Effective mass. a**, Longitudinal resistance $R_{xx}$ as a function of hole density under a 9 T magnetic field at various temperatures. **b**, $\Delta R_{xx}$ plotted as a function of $1/B$ for different temperatures at $n = -3.7\times10^{12}$ cm$^{-2}$, where the oscillatory component $\Delta R_{xx}$ is obtained by subtracting a linear background from the raw $R_{xx}$. Inset shows the temperature dependence of the $\Delta R_{xx}$ at $1/B = 0.1224$ T$^{-1}$. Red curve is the standard Lifshitz-Kosevich fit. **c**, The density dependence of the extracted effective mass $m^*$. For $|n| > 7\times10^{12}$ cm$^{-2}$, $m^*$ is extracted from the dataset shown in **a**. At lower densities, $m^*$ is determined from the measurements by scanning $B$ field at various temperatures, as shown in **b**. In the green-shaded region ($|n| < 3\times10^{12}$ cm$^{-2}$), $m^*$ is determined from Shubnikov-de Haas oscillations under spin/valley-polarized conditions, exhibiting a reduction compared to the values observed at higher densities. The dashed line denotes the average $m^* = 0.75\ m_e$ for $|n| > 3\times10^{12}$ cm$^{-2}$. All data in this figure are from Sample B.

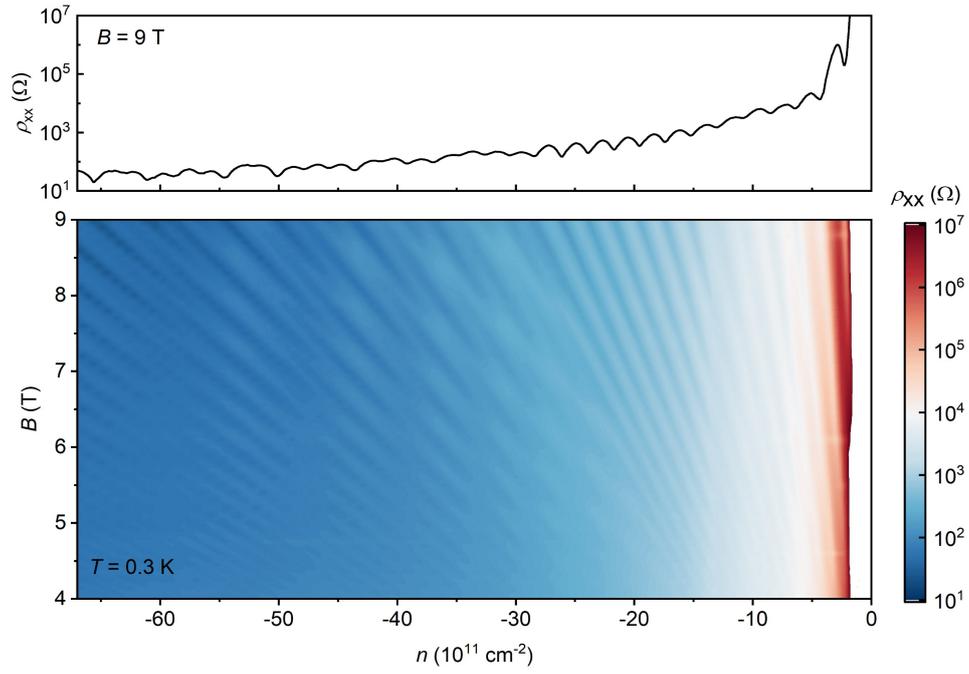

**Extended Data Figure 4 | Quantum oscillations in Sample A.** Mapping of $\rho_{xx}$ as a function of the magnetic field $B$ and the density $n$ at $T = 0.3$ K. The top panel shows a linecut at $B = 9$ T.

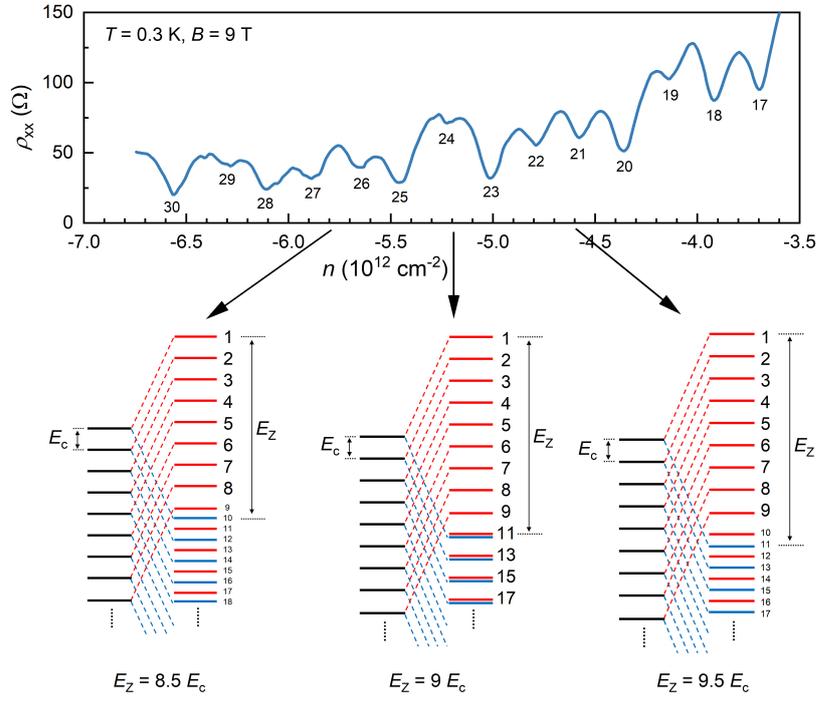

**Extended Data Figure 5 | Density dependent Zeeman energy.** The top panel shows the evolution of $\rho_{xx}$ (Sample A) within a narrow density range at $B = 9$ T and $T = 0.3$ K. Landau level filling factors $\nu_{LL}$ are labeled near the corresponding resistivity minima. Three schematics in the bottom panel depict Zeeman splitting of the Landau levels at three representative densities. Spin-up and spin-down Landau levels are denoted by red and blue lines, respectively. When $n \approx -5.8 \times 10^{12}$ cm$^{-2}$, $E_z = 8.5\, E_c$, the degeneracy of Landau levels is lifted evidently. When $n \approx -5.2 \times 10^{12}$ cm$^{-2}$, $E_z = 9\, E_c$, Landau levels exhibit nearly two-fold degeneracy for $\nu_{LL} \geq 11$. When $n \approx -4.6 \times 10^{12}$ cm$^{-2}$, $E_z = 9.5\, E_c$, the degeneracy lifting reemerges prominently.

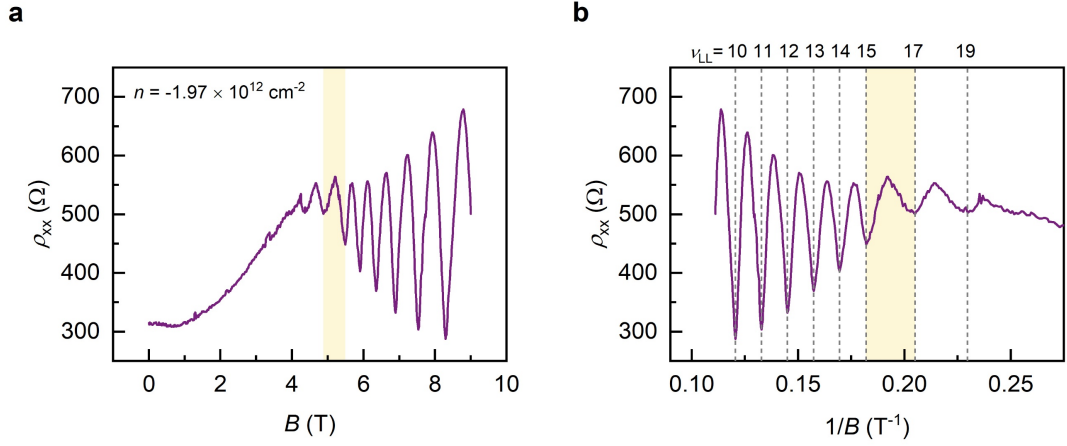

**Extended Data Figure 6 | Determination of $B_c$ from SdH Oscillations. a**, $\rho_{xx}$ (Sample A) as a function of magnetic field $B$ at $n = -1.97 \times 10^{12}$ cm$^{-2}$. **b**, Same $\rho_{xx}$ data as **a**, plotted versus inverse magnetic field $1/B$. Dashed lines highlight the change in oscillation period $\Delta(1/B)$. For $\nu_{LL} \leq 15$, the SdH oscillation frequency $B_F = (\Delta 1/B)^{-1} = 81.33$ T, yielding a Landau level degeneracy $D = \Phi_0|n|/B_F = 1$, where $\Delta(1/B)$ is the average oscillation period, $\Phi_0 = h/e$ is the magnetic flux quantum. For $\nu_{LL} > 15$, $B_F = 41.98$ T, $D = 2$. This implies a full spin/valley polarization at high fields ($\nu_{LL} \leq 15$), and a two-fold spin/valley degeneracy at low fields ($\nu_{LL} > 15$). The critical field $B_c$, marking the onset of full polarization, lies within the range of fields corresponding to $15 < \nu_{LL} < 17$ (yellow-shaded region), and is estimated from the median value.

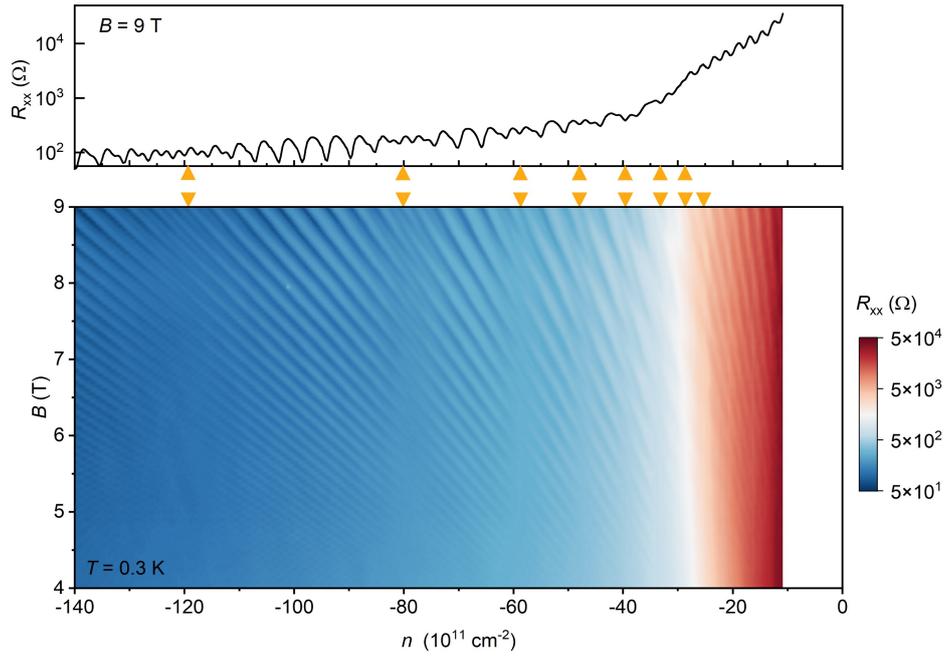

**Extended Data Figure 7 | Quantum oscillations in Sample B.** Mapping the dependence of $R_{xx}$ on magnetic field $B$ and the density $n$ at $T = 0.3$ K. The top panel shows a linecut at $B = 9$ T. The orange triangles mark the densities where spin/valley degeneracy is lifted evidently.

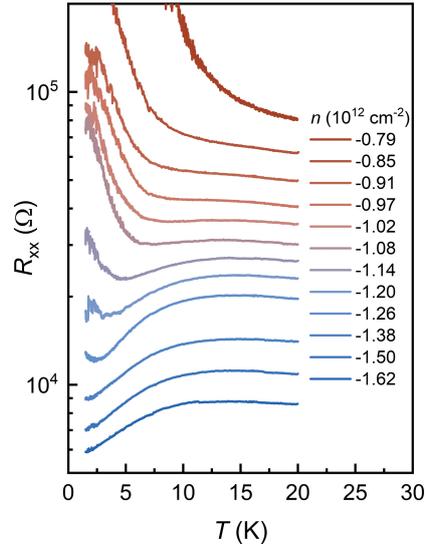

**Extended Data Figure 8 | MIT in Sample B.** Temperature dependence of the longitudinal resistance $R_{xx}$ at various hole densities, showing the density-tuned metal-insulator transition near a critical density $n_{MIT} \approx -1.1 \times 10^{12}$ cm$^{-2}$. Compared to sample A, the higher critical density observed in sample B is attributed to stronger disorder effects.